\documentclass[aps,pra,superscriptaddress,twocolumn, 10pt]{revtex4-1}
\usepackage{graphicx}
\usepackage{bm}
\usepackage[usenames]{color}
\bibstyle{apsrev.bib}

\usepackage{epsfig}
\usepackage{amsmath}

\usepackage{amssymb}
\usepackage{array}
\usepackage{nccmath}
\newcommand{\be}{\begin{equation}}
\newcommand{\ee}{\end{equation}}
\newcommand{\beqn}{\begin{eqnarray}}
\newcommand{\eeqn}{\end{eqnarray}}

\usepackage{dsfont}

\usepackage{array}

\begin{document}

\title{Entanglement of electrons and the lattice in a Luttinger system}

\author{Gerg\H{o} Ro\'osz}
\affiliation{Institute of Theoretical Physics, Technische Universit\"at Dresden, 01062 Dresden, Germany}
\affiliation{Institute for Solid State Physics and Optics, Wigner Research Centre for Physics, P.O. Box 49, 1525 Budapest, Hungary}

\author{Carsten Timm}
\email{carsten.timm@tu-dresden.de.}
\affiliation{Institute of Theoretical Physics, Technische Universit\"at Dresden, 01062 Dresden, Germany}
\affiliation{W\"urzburg-Dresden Cluster of Excellence ct.qmat, Technische Universit\"at Dresden, 01062 Dresden, Germany}

\date{September 7, 2020}

\begin{abstract}
The coupling between electronic and lattice degrees of freedom lies at the core of many important properties of solids. Nevertheless, surprisingly little is know about the entanglement between these degrees of freedom. We here calculate the entanglement entropy at zero temperature as well as the mutual information and the entanglement negativity at finite temperatures between the electrons and the lattice of a one-dimensional chain. The electrons are described within Luttinger-liquid theory. Our results show that the entanglement entropy diverges when one approaches the limit of stability, the so-called Wentzel-Bardeen singularity. We have found that the mutual information and the entanglement negativity decrease with the temperature. The mutual information reaches a finite value in the infinite-temperature limit, which is the consequence of the infinite linear electron spectrum of Luttinger theory. The entanglement negativity becomes exactly zero above a certain temperature, i.e., the lattice and the electrons become non-entangled above this temperature. If the electron-electron interaction is unscreened or weakly screened, this characteristic temperature diverges with the system size. However, if the interaction is strongly screened the characteristic temperature is finite and independent of the system size.  
\end{abstract}


\maketitle

\section{Introduction}
\label{sec:intro}

When studying solid states one usually starts with the adiabatic (and approximate) decoupling of the electron and lattice systems. However, these systems are not independent, and one can take their correlations into account by introducing an electron-phonon-coupling term to the Hamiltonian. The electron-phonon coupling causes many interesting phenomena in condensed matter system, for example BCS-type superconductivity \cite{maxwell1950, reynolds1950, allen1950, cooper1956, schrieffer1957, bardeen1957}, the Peierls instability \cite{peierls1955, lee1973, loss2010}, and charge-density-wave formation~\cite{gruner1988, vanyolos2006, miller2000}. The description of the coupled electron-phonon system is a nontrivial problem. Sophisticated approximations, such as diagrammatic perturbation theory \cite{midgal1958, eliashberg1960}, Monte-Carlo simulations \cite{mishchenko2014, prokofev1998, mishchenko2000}, and the tensor-network approach \cite{pino2018} have been developed.

Advances in quantum information theory in the last two decades have made it possible to quantify correlations and entanglement between (sub-) systems without depending on concrete correlation functions and observables \cite{amico2008, szalay2017}. This has led to a better understanding of thermalization \cite{serbyn2019} and simulability \cite{shi2006} of quantum systems. 
The entanglement entropy between the electron and the protons in the $H_2^+$ molecular ion has been calculated \cite{sanz-vicaro2017} but we are not aware of similar studies for extended systems. The present paper describes a step in this direction.

In order to obtain precise knowledge about the whole spectrum and the wave functions of all excited states, we use an integrable model, which on the other hand should be able to describe real systems. Such a model exists for one dimension, namely the Luttinger liquid coupled to acoustic phonons, which was introduced by Wentzel \cite{wentzel1950} and Bardeen \cite{bardeen1951}. This model is best known for the Wentzel-Bardeen singularity: For sufficiently strong electron-phonon coupling, the Hamiltonian becomes unbounded from below. Early work on the Wentzel-Bardeen singularity was motivated by its suspected analogy with superconductivity in higher-dimensional systems~\cite{wentzel1950, bardeen1951, varga1964}.

Similar models can be used to describe the electron-phonon coupling in carbon nanotubes \cite{suzuura2002, rosati2015, martino2003}, and there have been speculations that the Wentzel-Bardeen singularity could be realized in these systems \cite{martino2003}. Notably, in certain nanotube systems the electron-phonon interaction is tunable by adding quantum dots to the nanotube~\cite{benyamini2014}.

For the sake of a transparent and compact treatment, we here consider one electronic band and one acoustic phonon band. In order to describe nanotubes, one would have to add multiple electronic bands and several phonon branches. This is technically straightforward and can be the subject of later studies.

The correlations and entanglement will be characterized using the following measures: At zero temperature, the entanglement entropy is used to characterize the entanglement and the correlations \cite{bennett1996, eisert2010}. The system is in its ground state $|\mathrm{GS}\rangle$, and its density matrix is the projector $\rho = |\mathrm{GS}\rangle \langle\mathrm{GS}|$. One divides the system into two complementary parts $A$ and $B$, which in our case are the electrons and the lattice. The reduced density matrices of the two subsystems are 
\begin{equation}
\rho_A = \textrm{Tr}_B \rho , \quad \rho_B = \textrm{Tr}_A \rho ,
\label{eq:red_den_matrix}
\end{equation}
where $\mathrm{Tr}_A$ and $\mathrm{Tr}_B$ denote the partial trace over subsystem $A$ and $B$, respectively. The entanglement entropy is defined as the von Neumann entropy of the reduced density matrices,
\begin{equation}
S = - \textrm{Tr}_B \rho_B \ln \rho_B = - \textrm{Tr}_A \rho_A \ln \rho_A .
\end{equation}

At nonzero temperature, one can characterize the total (quantum and classical) correlations using the mutual information \cite{vedral2002}. To define the mutual information one first introduces the entropies of the reduced density matrices,
\begin{align}
S_A &= - \textrm{Tr}_A \rho_A \ln \rho_A , \\
S_B &= - \textrm{Tr}_B \rho_B \ln \rho_B .
\end{align}
These two entropies are generally different, $S_A \neq S_B$. The mutual information is defined as 
\begin{equation}
I(A:B) = S_A + S_B - S_{A \cup B} ,
\end{equation}
where $S_{A \cup B} = - \textrm{Tr} \rho \ln \rho$ is the entropy of the whole system. 


In order to characterize the \emph{quantum} correlations at nonzero temperature, we use the entanglement negativity \cite{sanpera97, sanpera98}. To define the negativity one requires the concept of the partial transpose $\rho^{T_A}$ of the density matrix, which is defined in terms of matrix elements with respect to the product basis of subsystems $A$ and $B$,
\begin{equation}
\langle a_i, b_j | \rho^{T_A} | a_n, b_m \rangle =  \langle a_n, b_j | \rho | a_i, b_m \rangle .
\end{equation} 
The partial transpose is unitarily equivalent to a time reversal in subsystem $A$.
It turns out that classical states, i.e., states without entanglement, have no knowledge about the common time direction. The partial transpose of the density matrix is then also a valid density matrix with all eigenvalues positive \cite{sanpera97, sanpera98}. However, if the state is entangled negative eigenvalues may occur in the partial transpose. The sum of these negative eigenvalues is a so-called entanglement monotone, i.e., it does not decrease in absolute value under local operations and classical communication (LOCC)~\cite{vidal2002}. 

The negativity is defined as the sum of the absolute values of the negative eigenvalues of the partial transpose,
\begin{equation}
{\cal N} = \sum_{\lambda_i<0} |\lambda_i|  = \frac{||\rho^{T_A} ||_1 -1 }{2} ,
\end{equation}
where the $\lambda_i$ are the eigenvalues of the partial transpose $\rho^{T_A}$ and $||\bullet||_1$ is the trace norm, which is defined as the sum of the absolute values of the eigenvalues. The logarithmic negativity is then defined as 
\begin{equation}
{\cal E}_{\cal N} = \ln (2 {\cal N}+1) = \ln ||\rho^{T_A} ||_1 .
\end{equation} 

The rest of this paper is organized as follows: In Sec.\ \ref{sec:model}, we define our model and present its solution. In Sec.\ \ref{sec:correlation_functions}, the values of the important correlation functions are given. In Sec.\ \ref{sec:measures}, we then express the entanglement measures in terms of integrals, which are evaluated numerically in Sec.\ \ref{sec:Results}. Finally, we summarize and discuss our results in Sec.~\ref{sec:conc}.

\section{Model}
\label{sec:model}

The lattice is modeled as a harmonic oscillator chain with periodic boundary conditions, which is coupled to a one-dimensional Luttinger liquid \cite{wentzel1950, bardeen1951, varga1964}. If one turns off the electron-electron interaction this model is  equivalent to the original Wentzel-Bardeen model studied in Refs.\ \cite{wentzel1950, bardeen1951, varga1964}. We note that this problem can be treated by integrating out the phonons, which gives an effective electronic model \cite{loss1995}. Here we do not follow this approach since we need to keep the phonons in order to characterize the electron-phonon entanglement and correlations. After bosonization, we use methods derived for oscillator systems \cite{audenaert2002, eisler2014} to characterize the entanglement.

The system is defined by the Hamiltonian 
\begin{align}
H &= -\sum_{\sigma=\pm 1/2} \int_{0}^{L} \frac{dx}{2\pi}\, v_f \Big[ \;^*_* \Psi_{\sigma, L}^{\dagger} (x)\,
   i \partial_x \Psi_{\sigma, L} (x) \;^*_* \nonumber \\ 
&\quad{}+ \;^*_* \Psi_{\sigma, R}^{\dagger} (x)\, i \partial_x \Psi_{\sigma, R} (x) \;^*_*  \Big]
  \nonumber \\
&{}+ \sum_{j=1}^N  \frac{p_j^2}{2} + \frac{1}{2}\, \kappa (q_j -q_{j+1})^2
  \nonumber \\
&{}+ \frac{1}{\sqrt{L}} \sum_{j=1}^L q_j \int_{0}^{L} dx\, \big[ \hat{n}_{L} (x)+\hat{n}_{R} (x) \big] g(|x- j|_L)
  \nonumber \\
&{}+ \frac{4}{L} \int_0^L \int_0^L dx\,dy\, \big(\hat{n}_L(x), \hat{n}_R(y)\big)
  \nonumber \\
&\quad{}\times \begin{pmatrix}
    h(x-y) &  \frac{1}{2}f(x-y) \\
    \frac{1}{2}f(x-y) & h(x-y)
  \end{pmatrix}
  \begin{pmatrix}
    \hat{n}_L(x) \\
    \hat{n}_R(y)
  \end{pmatrix}
\label{h-real-space}
\end{align}
where $L$ is the length of the system and the symbols $\;^*_* \bullet^*_*$ denotes normal ordering. The lattice constant is unity so that the number of the oscillators (atoms) is also $L$. The equilibrium positions of the atoms are $x_j = j$. The field operators $\Psi^{\dagger}_{\sigma, L} (x)$, $\Psi^{\dagger}_{\sigma, R} (x)$ create an electron with spin $\sigma=\pm 1/2$ at site $x$, and $L$, $R$ stand for the left-going and right-going electrons. The first two terms in Eq.\ (\ref{h-real-space}) denote the kinetic energy of the electrons, where the factor $2\pi$ stems from the normalization of the fields \cite{delft1998}. The third line describes the lattice system, where $q_i$ and $p_i$ are the selfadjoint canonical position and momentum operators of atom $i$. The fourth line corresponds to the electron-phonon coupling. The local electron densities are $\hat{n}_L (x)= \sum_{\sigma} \Psi_{\sigma, L}^{\dagger} (x) \Psi_{\sigma, L} (x)$ and $\hat{n}_R (x)= \sum_{\sigma} \Psi_{\sigma, R}^{\dagger} (x) \Psi_{\sigma, R} (x)$.    The translation of the oscillators couples to the electron density in a non-local manner described by the function $g(|x-j|_L)$, where $|x-j|_L$ is the shortest distance between $x$ and $j$, taking periodic boundary conditions into account. We do not consider umklapp processes. 


In the literature, it was supposed \cite{wentzel1950, bardeen1951, varga1964} that the Fourier transform $g_k$ of $g(\Delta x)$ is linear for small $k$, i.e., $g_k \sim k$. The origin of this assumption may be the Bloch formula which indeed predicts linear electron-phonon coupling \cite{bloch1929}. It has become clear, though, that the picture of a homogeneous positive background used in the derivation of the Bloch formula is too crude \cite{giustino2017} and that the electron-phonon coupling is generally not linear in the wave number. In a number of real one-dimensional systems, the electron-phonon coupling is found to be $g_k \sim \sqrt{k}$ \cite{suzuura2002, giustino2017}. We will discuss both forms of electron-phonon coupling below.

The last two lines in Eq.\ (\ref{h-real-space}) describe the electron-electron interaction. The interaction is assumed to be spin independent. The interaction between electrons moving in the same (opposite) directions is described by the function $h(x-y)$ ($f(x-y)$). Since the function $h(x-y)$ describes processes with small momentum transfer, whereas $f(x-y)$ corresponds to processes with large momentum transfer on the order of $2k_F$, $h(x-y)$ is expected to be larger than $f(x-y)$. Depending on screening, the Fourier transform $h_k$ of $h(x-y)$ may be singular at $k=0$. We will discuss singularities of the power-law form $h_k \sim 1/k^\alpha$ below. We suppose that the function $f(x-y)$ and its Fourier transform are regular. The factor of $4$ is included here for later convenience.

The electron operators in momentum space are
\begin{align}
c_{k,\sigma, \nu} &= \sqrt{\frac{2 \pi}{L}} \int_{-L/2}^{L/2} dx\, e^{ikx}\, \Psi_{\sigma, \nu} (x) , \\
c^{\dagger}_{k,\sigma, \nu} &= \sqrt{\frac{2 \pi}{L}} \int_{-L/2}^{L/2} dx\, e^{-ikx}\,
  \Psi^{\dagger}_{\sigma, \nu} (x) ,
\end{align}
where $k=(2\pi/L)\, n$ with $n \in \mathbb{Z}$. The factor $\sqrt{2 \pi}$ results from the normalization of the field \cite{delft1998}. In the next step, we construct the bosonic operators 
\begin{align}
b_{\sigma, \nu, q} &= \frac{1}{\sqrt{n_k}} \sum_{k=-\infty}^{\infty} c^{\dagger}_{\sigma, \nu, k+q}
  c_{\sigma, \nu,k} , \\
b^{\dagger}_{\sigma, \nu, q} &= \frac{1}{\sqrt{n_k}} \sum_{k=-\infty}^{\infty} c^{\dagger}_{\sigma, \nu,k}
  c_{\sigma, \nu, k+q} .
\end{align}
Using these bosonic operators, we define selfadjoint momentum and coordinate operators for the electronic degrees of freedom as
\begin{align}
q_{k,\eta, 1} &= \sum_{\sigma=\pm 1/2} \frac{(-1)^{\eta (\sigma+1/2)}}{\sqrt{8} \sqrt{\Omega_k}} \nonumber \\
&\quad{} \times \left( b^{\dagger}_{k,\sigma} + b_{k,\sigma} + b^{\dagger}_{-k,\sigma} + b_{-k,\sigma}
  \right) , \\
q_{k,\eta, 2} &= \sum_{\sigma=\pm 1/2 } \frac{-i(-1)^{\eta (\sigma+1/2)}}{\sqrt{8}  \sqrt{\Omega_k}} \nonumber \\
&\quad{} \times \left(  b^{\dagger}_{k,\sigma} - b_{k,\sigma}  - b^{\dagger}_{-k,\sigma} + b_{-k,\sigma}
  \right) , \\
p_{k,\eta, 1} &= \sum_{\sigma=\pm 1/2 } \frac{i \sqrt{\Omega_k} (-1)^{\eta (\sigma+1/2)}}{\sqrt{8}} \nonumber \\
&\quad{} \times \left(  b^{\dagger}_{k,\sigma} - b_{k,\sigma}  + b^{\dagger}_{-k,\sigma} - b_{-k,\sigma}
  \right) , \\
p_{k,\eta, 2} &= \sum_{\sigma=\pm 1/2 }  \frac{\sqrt{\Omega_k} (-1)^{\eta (\sigma+1/2)}}{\sqrt{8}} \nonumber \\
&\quad{} \times \left(  b^{\dagger}_{k,\sigma} + b_{k,\sigma} - b^{\dagger}_{-k,\sigma} - b_{-k,\sigma} \right) ,
\end{align}
where $k \ge 0$. Here, $\eta=0$ corresponds to the charge modes and $\eta=1$ corresponds to the spin modes. These operators satisfy the canonical commutation relations $[q_{k,\eta,j}, q_{q, \eta, i}] = [p_{k,\eta,j}, p_{q, \eta, i}]=0$ and $[q_{k, \eta, j}, p_{q, \eta', i}]=\delta_{q,k} \delta_{\eta, \eta'} \delta_{i,j}$. The frequency $\Omega_k$ is defined as $\Omega_k=v_f k + \frac{5}{4 \pi} h_k k$.

Turning to the lattice degrees of freedom, we introduce the Hermitian sine and cosine modes
\begin{align}
Q_{S,k} &= \sqrt{\frac{2}{L}} \sum_{n=1}^L \sin(kn)\, q_n , \\
P_{S,k} &= \sqrt{\frac{2}{L}} \sum_{n=1}^L \sin(kn)\, p_n , \\
Q_{C,k} &= \sqrt{\frac{2}{L}} \sum_{n=1}^L \cos(kn)\, q_n , \\
P_{C,k} &= \sqrt{\frac{2}{L}} \sum_{n=1}^L \cos(kn)\, p_n .
\end{align}
The inverse transformations read as
\begin{align}
q_n &= \sqrt{\frac{2}{L}} \sum_k \big[ \cos(kn)\, Q_{C,k} + \sin(kn)\, Q_{S,k} \big] , \\
p_n &= \sqrt{\frac{2}{L}} \sum_k \big[ \cos(kn)\, P_{C,k} + \sin(kn)\, P_{S,k} \big] .
\end{align}
We finally obtain the Hamiltonian in oscillator form,
\begin{align}
H &= \sum_{k=0}^\infty \sum_{\eta=0,1} \left( \Omega_k  + \frac{1}{2}\, p_{k, \eta, 1}^2
  + \frac{1}{2}\, \Omega_k^2 q_{k, \eta, 1}^2  \right. \nonumber \\
&\quad \left. {}+ \frac{1}{2}\, p_{k, \eta, 2}^2 + \frac{1}{2}\, \Omega_k^2 q_{k, \eta, 2}^2 \right) \nonumber \\
&{}+ \sum_{k=0}^{\pi} \frac{P_{S,k}^2 }{2m} + \frac{1}{2}\, \omega_k m Q_{S,k}^2 + \frac{P_{C,k}^2 }{2m}
  + \frac{1}{2}\, \omega_k m Q_{C,k}^2 \nonumber \\
&{}+ \sum_{k=0}^{\pi} g_k \sqrt{2 n_k \Omega_k} \left( Q_{C,k} q_{k,  0, 1} + Q_{S,k} q_{k, 0, 2} \right)
  \nonumber \\
&{}+ \sum_{k=0}^{\pi} \frac{5}{4\pi}\, k f_k \left( \frac{1}{2}\, \Omega_k q_{k,0,1}^2
  - \frac{1}{2\Omega_k}\, p_{k,0,1}^2 \right. \nonumber \\
&\quad \left. {}+ \frac{1}{2}\, \Omega_k q_{k,0,2}^2 - \frac{1}{2 \Omega_k}\, p_{k,0,2}^2 \right) ,
\label{h-osc-form}
\end{align}
where the frequency of the phonon modes is $\omega_k = 2 \sqrt{\kappa}\, |\sin k/2|$.

\subsection{Diagonalization}

The Hamiltonian in Eq.\ (\ref{h-osc-form}) can be diagonalized by a canonical transformation. The charge modes with $|k|> \pi$ do not couple to the lattice and are thus left unchanged during the diagonalization. Similarly, the spin modes, represented by $q_{k,1,1}$, $q_{k,1,2}$, $p_{k,1,1}$, $p_{k,1,2}$, do not couple to the lattice at all and are also unchanged.

 
The diagonalized Hamiltonian is  
\begin{align}
H &= \sum_{k=0}^{\infty} \left( \Omega_k  + \frac{1}{2}\, p_{k, 1, 1}^2 + \frac{1}{2}\, \Omega_k^2 q_{k, 1, 1}^2
  \right. \nonumber \\
&\quad \left. {}+ \frac{1}{2}\, p_{k, 1, 2}^2 + \frac{1}{2}\, \Omega_k^2 q_{k, 1, 2}^2 \right) \nonumber \\
&{}+ \sum_{k=\pi}^{\infty} \left( \frac{1}{2}\, p_{k, 0, 1}^2 + \frac{1}{2}\, \Omega_k^2 q_{k, 0, 1}^2 \right.
  \nonumber \\
&\quad \left. {}+ \frac{1}{2}\, p_{k, 0, 2}^2 + \frac{1}{2}\, \Omega_k^2 q_{k, 0, 2}^2 \right) \nonumber \\
&{}+ \sum_{k=0}^{\pi} \left( \frac{P_{1,+,k}^2 }{2} + \frac{1}{2}\, \lambda_{+,k} Q_{1,+,k}^2
  + \frac{P_{1,-,k}^2 }{2} \right. \nonumber \\
&\quad {}+ \frac{1}{2}\, \lambda_{-,k} Q_{1,+,k}^2 
  + \frac{P_{2,+,k}^2 }{2} + \frac{1}{2}\, \lambda_{+,k} Q_{2,-,k}^2 \nonumber \\
&\quad \left. {}+ \frac{P_{2,-,k}^2 }{2} + \frac{1}{2}\, \lambda_{-,k} Q_{2,+,k}^2 \right) ,
\end{align}
with
\begin{align}
\lambda_{k,\pm} &= \frac{1}{2} \left[ \omega_k^2  + \left(\Omega_k^2+ \frac{5}{4 \pi} k f_k \right) \alpha^2_k \right] \nonumber \\
&{}\pm \frac{1}{2} \sqrt{ \left[ \omega_k^2  - \left(\Omega_k^2+ \frac{5}{4 \pi} k f_k \right) \alpha^2_k \right]^2 + \frac{4}{ \pi} g_k^2 k \Omega_k \alpha^2_k} ,
\end{align}
where
\begin{equation}
\alpha_k = \sqrt{1-\frac{f_k}{\frac{4 \pi}{5} v_F + h_k} } \;.
\end{equation}
The nontrivial eigenfrequencies of the diagonalized Hamiltonian are given by $\sqrt{\lambda_{k,\pm}}$. The radicand $\lambda_{k,\pm}$ can be negative, in which case the Hamiltonian is not bounded from below and the system becomes unstable. This is known as the Wentzel-Bardeen singularity~\cite{wentzel1950, bardeen1951}.
%
%
The stability criterion reads as~\cite{varga1964}
\begin{equation}
\omega_k^2 \left[ v_f + \frac{5}{4 \pi}  (f_k +h_k)\right] > \frac{g_k^2}{\pi} .
\label{eq:stab_crit}
\end{equation}

The nontrivial eigenmodes diagonalizing the Hamiltonian are given by
\begin{align}
P_{1, \pm, k} &= \alpha_k \left( A_{\pm,k} P_{C,k} + B_{\pm,k} p_{k,0,1} \right) , \\
Q_{1, \pm, k} &= \frac{A_{\pm,k} Q_{C,k} + B_{\pm,k} q_{k,0,1}}{\alpha_k} , \\
P_{2, \pm, k} &= \alpha_k \left( A_{\pm,k} P_{S,k} + B_{\pm,k} p_{k,0,2} \right) , \\
Q_{2, \pm, k} &= \frac{A_{\pm,k} Q_{S,k} + B_{\pm,k} q_{k,0,2}}{\alpha_k} ,
\end{align}
where 
\begin{align}
A_{\pm,k} &= -\frac{1}{\sqrt{N_k}}\, g_k \sqrt{\frac{k \Omega_k}{4 \pi}}\,
  \sqrt{ 1- \frac{\frac{5}{4 \pi} f_k}{v_F + \frac{5}{4 \pi} h_k}} , \\
B_{\pm,k} &= \frac{1}{\sqrt{N_k}}\, (\omega^2_k - \lambda_{\pm}) , \\
N_k &= (\omega^2_k-\lambda_{\pm})^2
  + g^2_k \frac{k \Omega_k}{4 \pi} \sqrt{ 1- \frac{\frac{5}{4 \pi} f_k}{v_F + \frac{5}{4 \pi} h_k}} .
\end{align}
To complete the solution, we express these eigenmodes in terms of the bosonic operators
\begin{align}
a_{i,\pm,k} &= \frac{\lambda_{k,\pm}^{1/4}}{\sqrt{2}} \left( Q_{i,\pm,k}
  + i\, \frac{1}{\lambda^{1/4}_{k,\pm}}\, P_{i,\pm,k} \right) , \\
a^{\dagger}_{i,\pm,k} &= \frac{\lambda_{k,\pm}^{1/4}}{\sqrt{2}} \left( Q_{i,\pm,k}
  - i\, \frac{1}{\lambda^{1/4}_{k,\pm}}\, P_{i,\pm,k} \right) .
\end{align}
With the solution in hand, we can calculate the pair correlation functions.

\subsection{Correlation functions}
\label{sec:correlation_functions}

The entanglement measures can be calculated from pair correlation functions. The correlation functions of the lattice sine and cosine modes read as
\begin{align}
\langle Q^2_{S,k} \rangle &= \langle Q^2_{C,k} \rangle = \frac{1}{\alpha_k^2}
  \left[\frac{A^2_{k,+}}{2 \lambda^{1/2}_{k,+}}
  \left( \frac{2}{e^{\beta \lambda^{1/2}_{k,+}}-1 }+1 \right) \right. \nonumber \\
& \left. {}+ \frac{A^2_{k,-}}{2 \lambda^{1/2}_{k,-}}
  \left( \frac{2}{e^{\beta \lambda^{1/2}_{k,-}}-1 }+1 \right) \right] , \\
\langle P^2_{S,k} \rangle &= \langle P^2_{C,k} \rangle = \alpha_k^2
  \left[\frac{A^2_{k,+} \lambda^{1/2}_{k,+}}{2}
  \left( \frac{2}{e^{\beta \lambda^{1/2}_{k,+}}-1 }+1 \right) \right. \nonumber \\
& \left. {}+ \frac{A^2_{k,-} \lambda^{1/2}_{k,-}}{2 }
  \left( \frac{2}{e^{\beta \lambda^{1/2}_{k,-}}-1 }+1 \right) \right] .
\end{align}
The correlation functions of the charge modes of the electronic subsystem are
\begin{align}
\langle q_{k,0,1}^2 \rangle &= \langle q^2_{k,0,2} \rangle = \frac{1}{\alpha_k^2}
  \left[\frac{B^2_{k,+}}{2 \lambda^{1/2}_{k,+}} \left( \frac{2}{e^{\beta \lambda^{1/2}_{k,+}}-1 } +1 \right)
  \right. \nonumber \\
& \left. {}+ \frac{B^2_{k,-}}{2 \lambda^{1/2}_{k,-}} \left( \frac{2}{e^{\beta \lambda^{1/2}_{k,-}}-1 } +1 \right)
  \right] , \\
\langle p^2_{k,0,1} \rangle &= \langle p^2_{k,0,2} \rangle = \alpha_k^2
  \left[\frac{B^2_{k,+} \lambda^{1/4}_{k,+}}{2}
  \left( \frac{2}{e^{\beta \lambda^{1/4}_{k,+}}-1 } +1 \right) \right. \nonumber \\
& \left. {}+ \frac{B^2_{k,-} \lambda^{1/4}_{k,-}}{2}
  \left( \frac{2}{e^{\beta \lambda^{1/4}_{k,-}}-1 } +1 \right) \right] .
\end{align}
Finally, the correlation functions connecting the lattice and electronic charge modes read as
\begin{align}
\langle Q_{S,k}\, q_{k,0,2} \rangle &= \langle Q_{C,k}\, q_{k,0,1} \rangle \nonumber \\ 
&= \frac{A_{+,k} B_{+,k}}{2 \lambda^{1/2}_{k,+}}
  \left( \frac{2}{e^{\beta \lambda^{1/2}_{k,+}}-1 } +1 \right) \nonumber \\
&{}+ \frac{A_{-,k} B_{-,k}}{2 \lambda^{1/2}_{k,-}}
  \left( \frac{2}{e^{\beta \lambda^{1/2}_{k,-}}-1 } +1 \right) , \\
\langle P_{S,k}\, p_{k,0,2} \rangle &= \langle P_{C,k}\, p_{k,0,1} \rangle \nonumber \\
&= \frac{A_{+,k} B_{+,k} \lambda^{1/2}_{k,+}}{2}
  \left( \frac{2}{e^{\beta \lambda^{1/2}_{k,+}}-1 } +1 \right) \nonumber \\
&{}+ \frac{A_{-,k} B_{-,k} \lambda^{1/2}_{k,-}}{2}
  \left( \frac{2}{e^{\beta \lambda^{1/2}_{k,-}}-1 } +1 \right) .
\end{align}
The correlation functions connecting the electronic spin modes to the charge or lattice modes are zero.

\section{Entanglement and correlation measures} 
\label{sec:measures}

In this section we describe the calculation of the measures used to quantify entanglement and correlations. Most of them are expressed by sums of analytical terms. The starting point is the equilibrium density matrix of the system, which is a bosonic Gaussian operator.
   
\subsection{Entanglement entropy}

At zero temperature, we consider the entanglement entropy between the electron and lattice degrees of freedom. The entanglement entropy is calculated using the correlation-function method \cite{peschel2012}, which has been used to investigate momentum-space entanglement in a Luttinger liquid \cite{dora2017}. One first defines the correlation matrices of the lattice,
\begin{equation}
Q_{i,j} = \langle\mathrm{GS}| q_i q_j|\mathrm{GS}\rangle, \quad
P_{i,j} = \langle\mathrm{GS}| p_i p_j|\mathrm{GS}\rangle .
\end{equation}
Let the spectrum of the matrix $C=QP$ be $\nu_1 \dots \nu_L$. The entanglement entropy is then
\begin{align}
S &= \sum_{j=1}^L \left[ \left(\sqrt{\nu_j} + \frac{1}{2} \right) \ln \left(\sqrt{\nu_j} + \frac{1}{2}\right)
  \right. \nonumber \\
& \left. {}- \left(\sqrt{\nu_j}-\frac{1}{2}\right) \ln \left(\sqrt{\nu_j}-\frac{1}{2}\right) \right] .
\end{align}
Introducing the function $s(x) = (\sqrt{x} + 1/2)\ln (\sqrt{x} + 1/2) -(\sqrt{x}-1/2)\ln (\sqrt{x}-1/2)$, one can rewrite this as
\begin{equation}
S = \mathrm{Tr}\, s(C) , 
\end{equation}
where the trace has to be computed on the $L$ dimensional space of $C$. Noting that $Q$ and $P$ have a common eigenbasis, namely the sine-cosine basis, and performing the trace with respect to this basis, we obtain a simple equation for the entanglement entropy:
\begin{equation}
S = \sum_{k>0}^{\pi} \left[ s(\langle Q^2_{S,k} \rangle \langle P^2_{S,k} \rangle)
  + s(\langle Q^2_{C,k} \rangle \langle P^2_{C,k} \rangle) \right] .
\end{equation}
This sum is easily calculated numerically. The results are shown in Sec.\ \ref{sec:Results}. In the thermodynamic limit, $L \gg 1$, the sum is replaced by the integral
\begin{equation}
S = \frac{L}{ 2 \pi} \int_0^{\pi} dk \left[ s(\langle Q^2_{S,k} \rangle \langle P^2_{S,k} \rangle)
  + s(\langle Q^2_{C,k} \rangle \langle P^2_{C,k} \rangle) \right] .
\label{eq:entl-ent-int}
\end{equation}
Since the integral is finite the entropy scales as $S \sim L$.

The reduced density matrix $\rho_\mathrm{ph}$ of the phonons can be written in a simple form using the selfadjoint canonical operators of the sine and cosine modes. For any temperature including $T=0$, it reads as
\begin{align}
\rho_\mathrm{ph} &= \frac{1}{Z} \prod_{k>0}^{\pi} e^{-\beta^\mathrm{eff}_{k} (\frac{1}{2} P^2_{C,k}
  + \frac{1}{2} \omega^\mathrm{eff}_k Q^2_{C,k})} \nonumber \\
& {}\times e^{-\beta^\mathrm{eff}_{k} (\frac{1}{2} P^2_{S,k}  + \frac{1}{2} \omega^\mathrm{eff}_k Q^2_{S,k})} ,
\end{align}
where
\begin{align}
\beta_k^\mathrm{eff} &= \frac{\sigma^Q_k}{\sigma^P_k}
  \ln \frac{\sigma^Q_k \sigma^P_k + 1/2}{\sigma^Q_k \sigma^P_k - 1/2} , \\
\omega^\mathrm{eff}_k &= \frac{\sigma^P_k}{\sigma^Q_k}
\end{align}
and $\sigma^Q_k$ and $\sigma^P_k$ are the variances of the coordinate and momentum operators
\begin{align}
\sigma^Q_k &= \sqrt{\langle Q^2_{C,k} \rangle}  = \sqrt{\langle Q^2_{S,k} \rangle} , \\
\sigma^P_k &= \sqrt{\langle P^2_{C,k} \rangle}  = \sqrt{\langle P^2_{S,k} \rangle} ,
\end{align}
respectively.
 
\subsection{Mutual information}

At nonzero temperatures, we calculate the mutual information between the electron and lattice degrees of freedom. It characterize the total correlation between the two subsystems. The mutual information is defined as
\begin{equation}
I = S_\mathrm{ph}+S_\mathrm{el}-S_{\mathrm{ph} \cup \mathrm{el}} .
\label{eq:I}
\end{equation}
The density matrix of the whole system can be written as
\begin{equation}
\rho_{\mathrm{ph} \cup \mathrm{el}} = \rho_1 \rho_s \rho_c ,
\end{equation}
with
%
\begin{align}
\rho_1 &= \frac{1}{Z_1} \prod_{|q|>\pi, \sigma}^{\infty} e^{- \beta |q| v_f b^\dagger_{q,\sigma} b_{q,\sigma}} ,
  \\ 
\rho_s &= \frac{1}{Z_s} \prod_{|q|<\pi} e^{-\beta |q| v_f b^\dagger_{S,q,\sigma} b_{S,q,\sigma}} , \\
\rho_c &= \frac{1}{Z_c} \prod_{k>0,\pm,j=1,2}^{\pi} e^{-\beta \lambda^{1/2}_{k,\pm} a^{\dagger}_{k,\pm,j}
  a_{k,\pm,j}} .
\end{align}
Here, $\rho_1$ describes the uncoupled short-wavelength excitations, which are present because the fermions are described by a continuum model, $\rho_s$ describes the uncoupled electronic spin modes, and $\rho_{c}$ describes the coupled electron-phonon modes. $Z_1$, $Z_s$, and $Z_c$ are the corresponding partition functions, which simply ensure that the density matrices have unit trace.

The terms from the uncoupled electronic modes cancel each other in Eq.\ (\ref{eq:I}) and the mutual information is determined only by the density matrix of the coupled modes,
\begin{equation}
I = S(\textrm{Tr}_\mathrm{ph} \rho_{c}) + S(\textrm{Tr}_\mathrm{el} \rho_{c}) - S(\rho_{c}) .
\end{equation}
The three terms are given by
\begin{align}
S(\textrm{Tr}_\mathrm{ph} \rho_{c}) &= \sum_{k>0}^{\pi}
  \left[ s(\langle Q^2_{S,k} \rangle \langle P^2_{S,k} \rangle)
  + s(\langle Q^2_{C,k} \rangle \langle P^2_{C,k} \rangle) \right] , \\
S(\textrm{Tr}_\mathrm{el} \rho_{c}) &= \sum_{k>0}^{\pi}
  \left[ s(\langle q^2_{k,0,1} \rangle \langle p^2_{k,0,1} \rangle)
  + s(\langle q^2_{k,0,2} \rangle \langle p^2_{k,0,2} \rangle) \right] , \\
S(\rho_{c}) &= 2 \sum_{k>0, \pm}^\pi \bigg[ \frac{\sqrt{\lambda}_{k,\pm}}{\exp(\beta \sqrt{\lambda}_{k,\pm}) -1 }
  \nonumber \\
&{}- \ln\left(1- \exp(-\beta \sqrt{\lambda}_{k,\pm})\right) \bigg] .
\label{eq:mut_inf}
\end{align}
These sums are easily calculated numerically. The results are presented in Sec.\ \ref{sec:Results} below


\subsection{Entanglement negativity}

As noted in Sec.\ \ref{sec:intro}, the logarithmic negativity is defined as
\begin{equation}
{\cal E}_{\cal N} = \ln (2 {\cal N}+1) = \ln ||\rho^{T_A} ||_1 .
\end{equation}
The partial transpose can be considered for any factor space of the Hilbert space. Here we would like to characterize the electron-phonon entanglement, therefore we consider the partial transpose for the phonon sector. The density matrix of the system is the tensor product
\begin{equation}
\rho_{\mathrm{ph} \cup \mathrm{el}} = \rho_1 \otimes \rho_s \otimes \rho_{c} .
\end{equation}
The partial transpose only affects the third term. We write the result as
\begin{equation}
\rho^{\Gamma}_{\mathrm{ph} \cup \mathrm{el}} = \rho_1 \otimes \rho_s \otimes \rho^{T_\mathrm{ph}}_{c} \;.
\end{equation}
We introduce the following notation for the eigenvectors and eigenvalues of the operators $\rho_1$, $\rho_s$, and $ \rho^{T_\mathrm{ph}}_{c}$:
\begin{align}
\rho_1 v_i &= a_i v_i    \\ 
\rho_s u_j &= b_j u_j  \\
\rho_{c}^{T_\mathrm{ph}} w_k &= c_k w_k .
\end{align}
Here, $a_i, b_j \in [0,1]$, whereas $c_k$ can be negative. The eigenvalues of the partially transposed full density matrix $\rho_{\mathrm{ph} \cup \mathrm{el}}^{T_\mathrm{ph}}$ are $\lambda_{i,j,k} = a_i b_j c_k$. Then the negativity reads as 
\begin{align}
{\cal{N}} &= \sum_{\lambda_{i,j,k}<0} |\lambda_{i,j,k}| \nonumber \\
&= \underbrace{\sum_i a_i}_{=\:1} \underbrace{\sum_j b_j}_{=\:1} \sum_{k\:\mathrm{with}\:c_k<0} |c_k|
  = \sum_{k\:\mathrm{with}\:c_k<0} |c_k| .
\end{align}
The sum of the negative eigenvalues of the full, partially transposed density matrix $\rho_{\mathrm{ph} \cup \mathrm{el}}^{T_\mathrm{ph}}$ is equal to the sum of the negative eigenvalues of $\rho^{T_\mathrm{ph}}_{c}$. Hence, we obtain the negativity by investigating only $\rho^{T_\mathrm{ph}}_{c}$. Since this is a Gaussian density matrix, we can determine the entanglement negativity from its covariance matrix, which has dimension $4N\times 4N$ \cite{audenaert2002, eisler2014}. To define the covariance matrix, one considers all possible expectation values of coordinate and momentum products. The real parts of these products gives the elements of the covariance matrix, for any arbitrary but fixed ordering of the operators. One can order the rows and columns of the covariance matrix in such a way that it is block diagonal. It then takes the form
\begin{widetext}
\begin{equation}
\begin{pmatrix}
\operatorname{Re} \langle Q_{k,c} Q_{k,c} \rangle & \operatorname{Re}\langle Q_{k,c} q_{k,0,1} \rangle &
  0 & 0 \\
\operatorname{Re} \langle Q_{k,c} q_{k,0,1} \rangle & \operatorname{Re}\langle q_{k,0,1} q_{k,0,1}  \rangle &
  0 & 0 \\
0 & 0 &
  \operatorname{Re}\langle P_{k,c} P_{k,c} \rangle & \operatorname{Re}\langle P_{k,c} p_{k,0,1} \rangle \\
0 & 0 &
  \operatorname{Re}\langle P_{k,c} p_{k,0,1} \rangle & \operatorname{Re}\langle p_{k,0,1} p_{k,0,1} \rangle
\end{pmatrix}
\label{diag1}
\end{equation}
for the cosine modes and
\begin{equation}
\begin{pmatrix}
\operatorname{Re} \langle Q_{k,s} Q_{k,s} \rangle & \operatorname{Re}\langle Q_{k,s} q_{k,0,2} \rangle &
  0 & 0 \\
\operatorname{Re}\langle Q_{k,s} q_{k,0,2} \rangle & \operatorname{Re}\langle q_{k,0,2} q_{k,0,2}  \rangle &
  0 & 0 \\
0 & 0 &
  \operatorname{Re}\langle P_{k,s} P_{k,s} \rangle & \operatorname{Re}\langle P_{k,s} p_{k,0,2} \rangle \\
0 & 0 &
  \operatorname{Re}\langle P_{k,s} p_{k,0,2} \rangle & \operatorname{Re}\langle p_{k,0,2} p_{k,0,2} \rangle
\end{pmatrix}
\label{diag2}
\end{equation}
\end{widetext}
for the sine modes. To get the covariance matrix of the partial transpose, one has to multiply every $P_{k,s}$ and $P_{k,c}$ with $-1$ in Eqs.\ (\ref{diag1}) and (\ref{diag2}).

One then obtains the logarithmic negativity from the symplectic eigenvalues of the covariance matrix of the partial transpose,
\begin{equation}
{\cal N} = - \sum_{\lambda} \ln \textnormal{min}(1, \lambda) ,
\end{equation}
where the sum runs over all symplectic eigenvalues. For our case, we find
\begin{equation}
{\cal N} = - 4 \sum_{k,\pm} \ln \textnormal{min}(1, \sqrt{\Lambda_{k, \pm} }) ,
\end{equation}
where
\begin{equation}
\Lambda_{k, \pm} = \frac{1}{2} \left[a_k \pm  \sqrt{a_k^2 +4 b_k -4 c_k} \right] ,
\end{equation}
with
\begin{align}
a_k &= \langle Q_{k,s} Q_{k,s} \rangle \langle P_{k,s} P_{k,s} \rangle
  + \langle q_{k,0,2} q_{k,0,2} \rangle \langle p_{k,0,2} p_{k,0,2} \rangle \nonumber \\
&{}+ 2 \langle P_{k,s} p_{k,0,2} \rangle  \langle Q_{k,s} q_{k,0,2} \rangle , \\
b_k &= \left( \langle Q_{k,s} q_{k,0,2} \rangle \langle P_{k,s} P_{k,s} \rangle
  - \langle q_{k,0,2} q_{k,0,2} \rangle \langle P_{k,s} p_{k,0,2} \rangle  \right) \nonumber \\
&{} \times \left(\langle Q_{k,s} q_{k,0,2} \rangle  \langle p_{k,0,2} p_{k,0,2} \rangle
  - \langle P_{k,s} p_{k,0,2} \rangle   \langle Q_{k,s} Q_{k,s} \rangle \right) , \\
c_k &= \left(\langle Q_{k,s} Q_{k,s} \rangle \langle P_{k,s} P_{k,s} \rangle
  - \langle P_{k,s} p_{k,0,2} \rangle  \langle Q_{k,s} q_{k,0,2} \rangle \right) \nonumber \\
&{} \times \left(\langle q_{k,0,2} q_{k,0,2} \rangle \langle p_{k,0,2} p_{k,0,2} \rangle -
      \langle P_{k,s} p_{k,0,2} \rangle  \langle Q_{k,s} q_{k,0,2} \rangle \right) .
\label{eq:ent_res}
\end{align}
In the general case, a positive value of the logarithmic negativity implies violation of separability but a zero value does not imply separability. However, our model consists of pairs of mutually coupled effective harmonic oscillators. It has been shown in Ref.\ \cite{werner2001} that in this case zero logarithmic negativity is equivalent to separability.    


\section{Results}
\label{sec:Results}

In this section, we evaluate the expressions derived in Sec.\ \ref{sec:measures} for two set of parameters. The first set corresponds to the original Wentzel-Bardeen model without electron-electron interaction and linear electron-phonon coupling, $g_k \sim k$, whereas the second describes a more realistic setting including electron-electron interaction and electron-phonon coupling $g_k \sim \sqrt{k}$ interaction matrix element, and with non-zero electron-electron interaction.


\subsection{Non-interacting model}

In this subsection, we investigate the original form of the Wentzel-Bardeen model, with vanishing interactions and electron-phonon coupling $g_k = g_0 k$. With these parameters, the stability criterion becomes
\begin{equation}
2 \omega^2_k \Omega_k > g^2_k n_k ,
\label{eq:non-int-stab} 
\end{equation}
which agrees with the results of Refs.\ \cite{wentzel1950, bardeen1951}. If one would consider $g_k \sim \sqrt{k}$, which we do not do in this subsection, the non-interacting model would be unstable for \emph{every} coupling strength since the left-hand side of Eq.\ (\ref{eq:non-int-stab}) scales with $\sim k^3$, and the right hand side would then scale with $\sim k^2$. We return to this point in the following subsection. 

For a sine-shaped dispersion of the phonons, first the highest-energy $k=\pi$ mode becomes unstabl, and the stable region is given by
\begin{equation}
v_f > \frac{\pi}{16} \frac{g_0^2}{\kappa} .
\end{equation}    
In the literature, there was a series of investigations to clarify the physical nature of this singularity. In our point of view, we use here a simple model without any anharmonic terms, which is only physical for a subset of the possible parameters. If the lattice is unstable in this model the only physical consequence is that in the corresponding regime the anharmonic terms cannot be neglected.

\begin{figure}
\includegraphics[width=8.3cm]{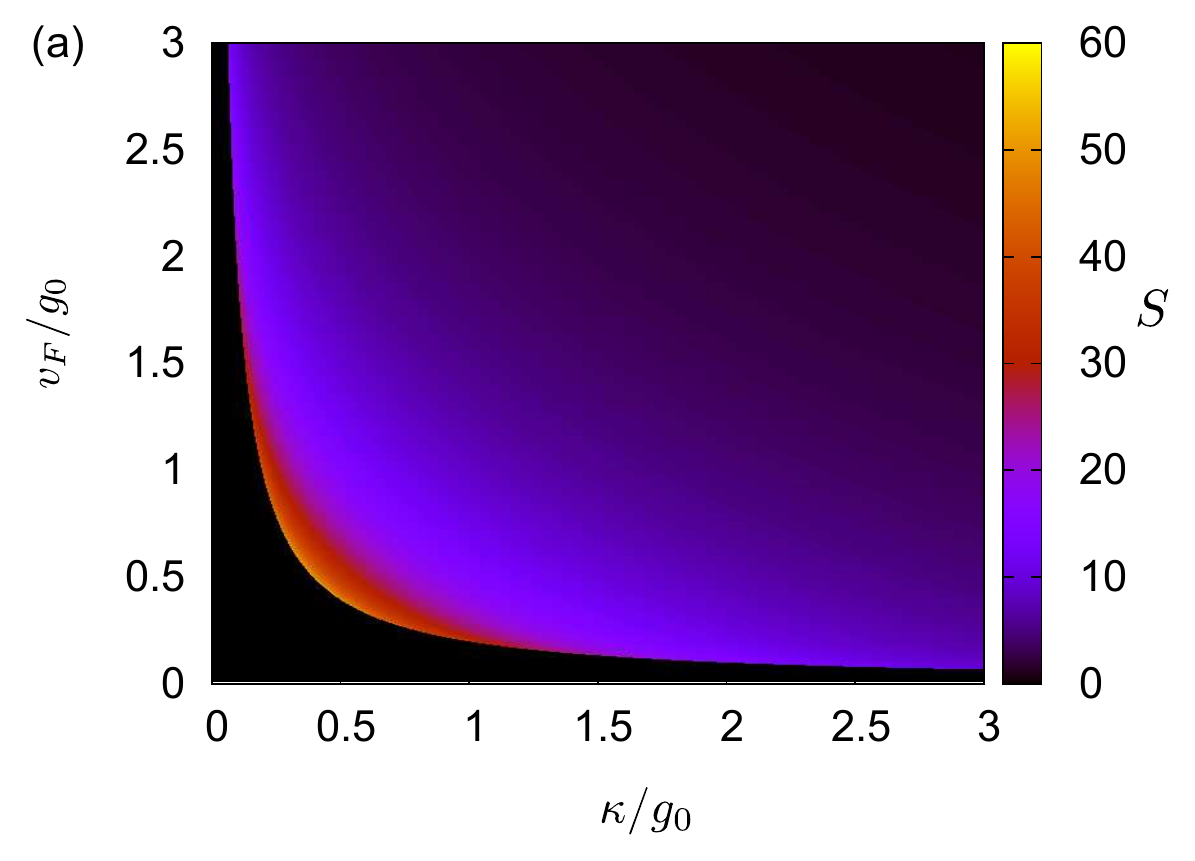}
\includegraphics[width=8.3cm]{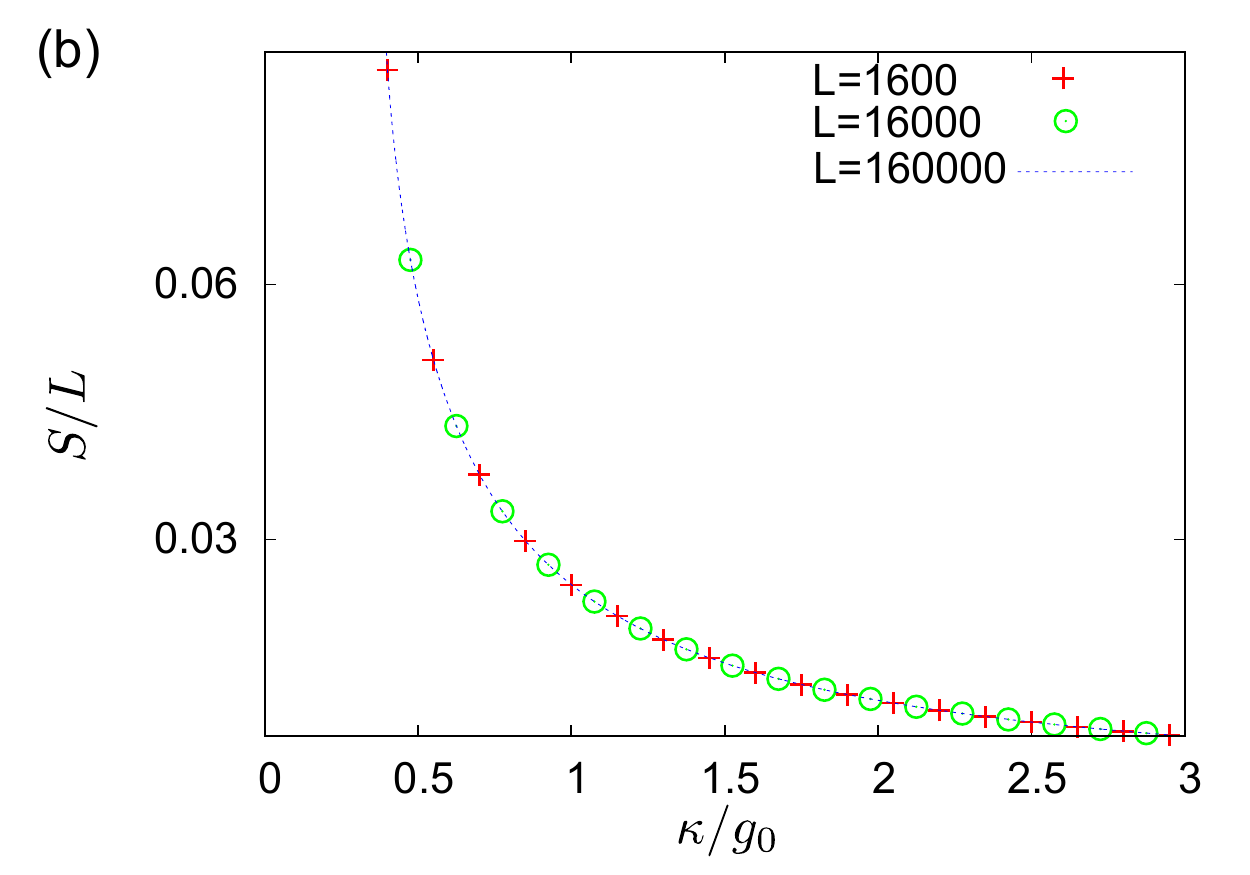}
\caption{Ground-state entanglement for the original non-interacting Wentzel-Bardeen model. (a) Entanglement entropy $S$ between the lattice and the electrons for a chain of length $L=200$, as a function of the lattice stiffness constant $\kappa$ and the Fermi velocity $v_f$. In the black regions close to the axes the system is unstable ($v_f < \pi/16\kappa$). (b) Entanglement entropy per length as a function of the stiffness $\kappa$ for $v_f=g_0/2$ for various system sizes.
\label{fig:ent_map}}
\end{figure}

\begin{figure}
\includegraphics[width=8.1cm]{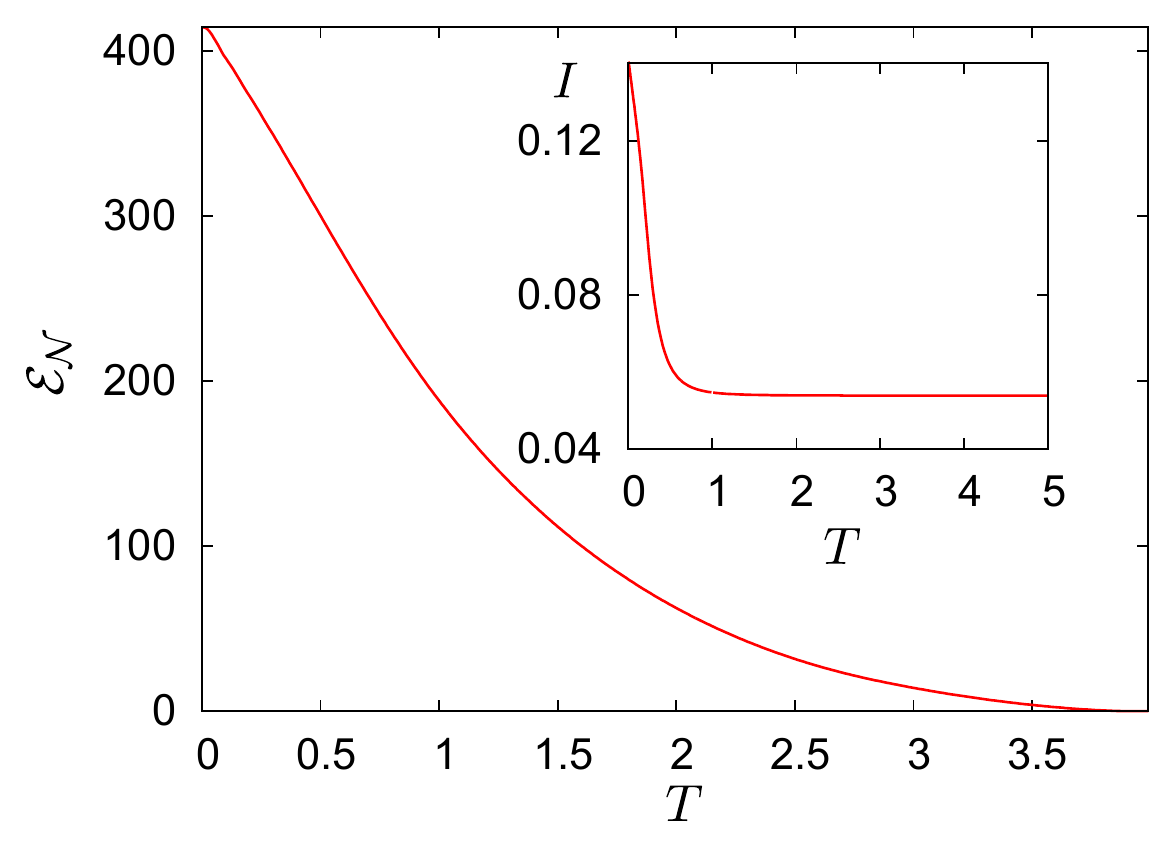}
\caption{Logarithmic negativity $\mathcal{E}_\mathcal{N}$ as a function of temperature for $L=100$, $\kappa=v_f=1$, and $g_0=0.1$. Inset: Mutual information of the non-interacting Wentzel-Bardeen model as a function of temperature  for the same parameters.
\label{fig:mutual_information_neg}}
\end{figure}

The entanglement entropy at temperature $T=0$ is shown as the function of the lattice stiffness and the Fermi velocity in Fig.\ \ref{fig:ent_map}. The entropy diverges close to the Wentzel-Bardeen singularity and is proportional to the system size, $S \sim L$.

The entanglement negativity is plotted in Fig.\ \ref{fig:mutual_information_neg}. The negativity decreases with increasing temperature and becomes exactly zero at and above a certain temperature. Similar behavior was observed in Ref.\ \cite{audenaert2002} for the entanglement negativity of a bisectioned harmonic chain. The mutual information is shown in the inset of Fig.\ \ref{fig:mutual_information_neg}. It decreases with increasing temperature and for high temperatures approaches a nonzero constant. This nonzero high-temperature value is a consequence of the infinite bandwidth of the model. In a model with a finite bandwidth, the bandwidth sets a temperature scale, and one expects that the mutual information exponentially falls to zero above that scale.  

\subsection{Interacting Wentzel-Bardeen model}
\label{sec:interacting_BW_model}

In this subsection, we consider a nonzero electron-electron interaction and electron-phonon coupling of the form $g_k = g_0 \sqrt{k}$. As noted above, for the non-interacting model this type of coupling always causes an instability. We assume that the interaction between electrons moving in the same direction is not too strongly screened, i.e, that it shows a singularity of the form
\begin{equation}
h_k = \frac{h_0}{|k|^{1+\alpha}} 
\end{equation}
for small $k$.
The interaction between electrons moving in opposite direction is assumed to show the same functional relationship but shifted by the momentum transfer $2k_F$ between right-moving the left-moving electrons at the Fermi energy, i.e.,
\begin{equation}
f_k = \frac{f_0}{(|k|+2 k_F)^{1+\alpha}} .
\end{equation}
Since $f_k$ is nearly constant for low-energy modes, its specific form should not affect the qualitative results.

Under these assumptions, the left-hand side of the stability equation is proportional to $k^{2+2\alpha}$, while the right-hand side is proportional to $k^{2+\alpha}$. We conclude that for $\alpha \ge 0$ the system can be stable. Detailed stability investigation can be performed by plotting the two sides of the stability criterion, Eq.~(\ref{eq:stab_crit}).

\begin{figure}
\includegraphics[width=8.3cm]{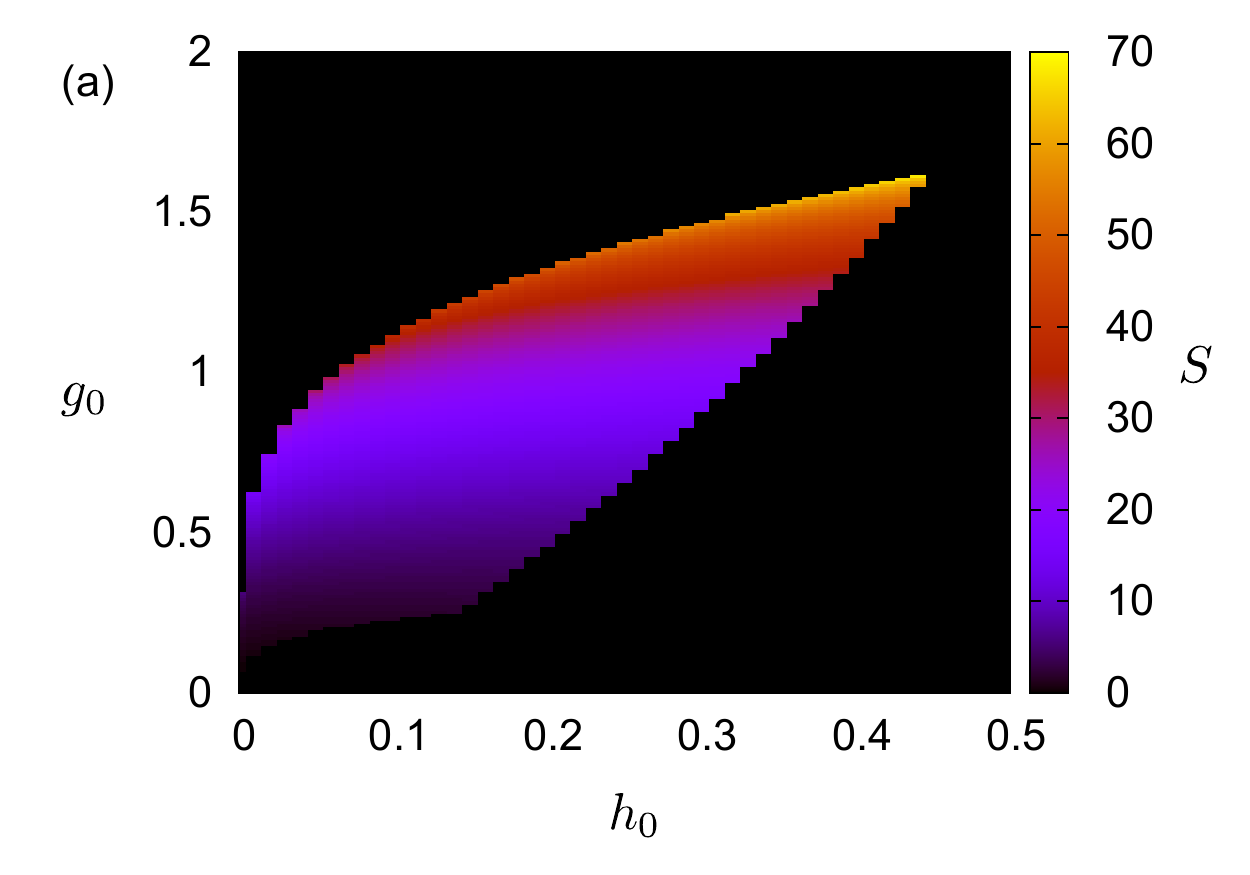}
\includegraphics[width=8.3cm]{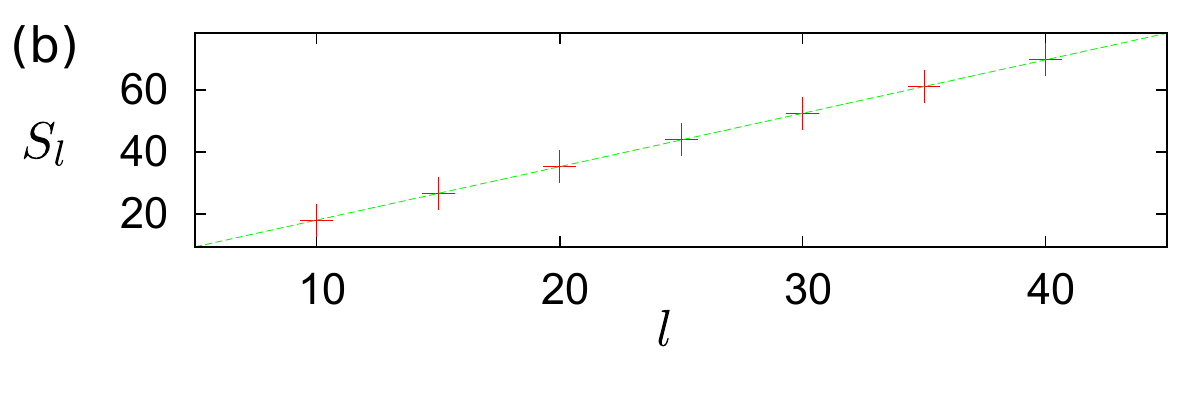}
\caption{(a) Entanglement entropy  $S$, for the interacting Wentzel-Bardeen model as a function of the electron-electron interaction parameter $h_0$ and the electron-phonon coupling parameter $g_0$. The Fermi velocity and the lattice stiffness are taken to be $v_f=\kappa=1$. The black region denotes that the system is unstable. (b) Entanglement entropy of a finite interval of the lattice of length $l<L$ and the rest of the system, including the other part of the lattice and all electronic degrees of freedom.
\label{fig:int_ent}}
\end{figure}

Results for the zero-temperature entanglement entropy of the interacting model are shown in Fig.\ \ref{fig:int_ent}(a). The entanglement increases with the electron-phonon coupling constant $g_0$, and decreases with the electron-electron interaction parameter $h_0$. Too large $h_0$ or too large $g_0$ render the system unstable. The entanglement entropy diverges when $g_0$ approaches the stability limit but remains finite if the stability limit is reached by increasing~$h_0$. 

It is also of interest to check how the entanglement entropy of a finite region scales with its size. It is known that the entanglement entropy of a coupled oscillator system defined on a lattice follows an area law, i.e., the entanglement entropy between two subsystems scales with the size of the surface dividing the subsystems, which for a chain is a point, scaling with $L^0$. On the other hand, the entanglement entropy of a fermionic system generally follows an area law with log corrections.
We have calculated the entanglement entropy between a finite part of the lattice of length $l<L$ and the rest of the system, consisting of the rest of the lattice and all electronic degrees of freedom, see Fig.\ \ref{fig:int_ent}(b). We evidently find volume-law scaling, $S_l\sim l$. The entanglement entropy between the whole lattice and the electrons is proportional to the full system size, $S \sim L$. We suggest that a similar phenomenon may occur for trapped cold atoms in an optical resonator, where the photons play the role of the coupled bosons.


\begin{figure}
\includegraphics[width=8.3cm]{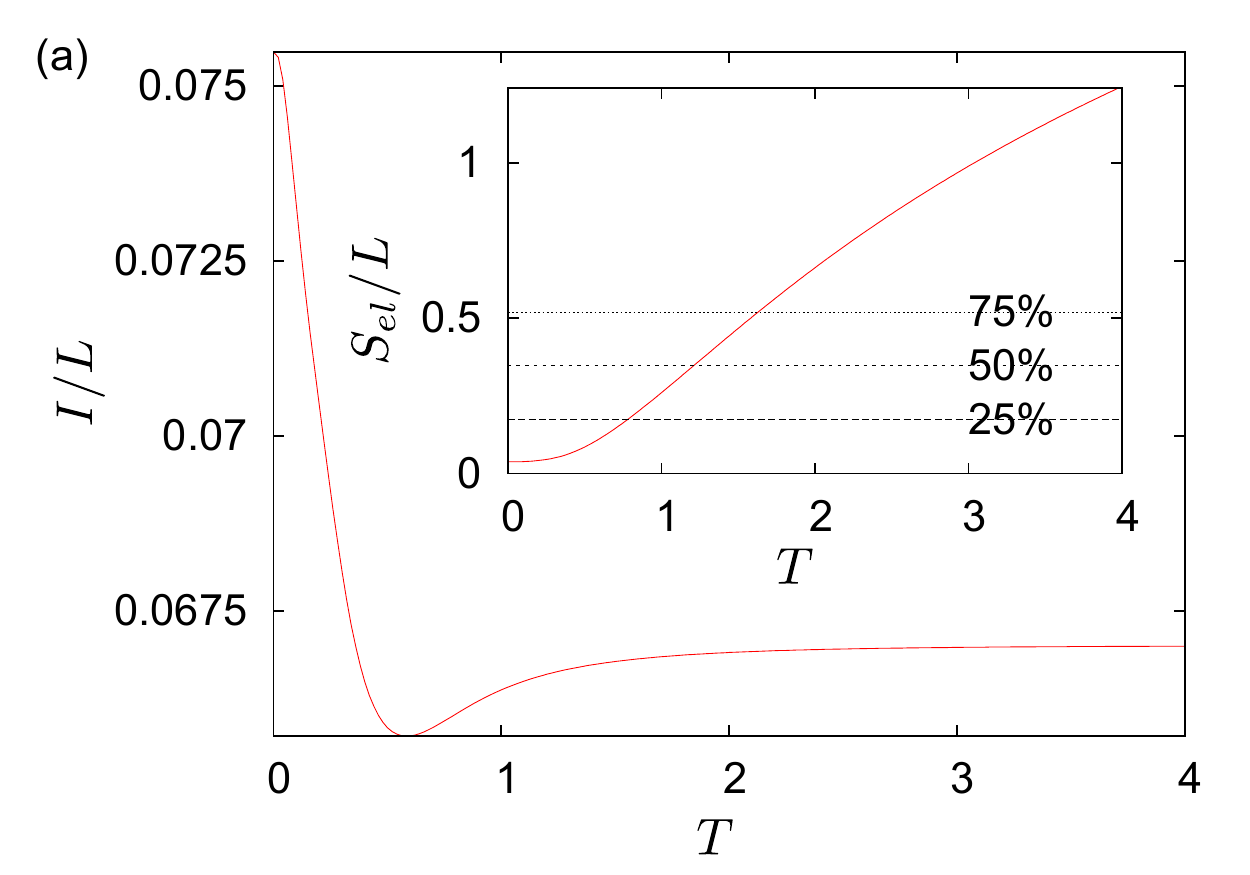}
\includegraphics[width=8.3cm]{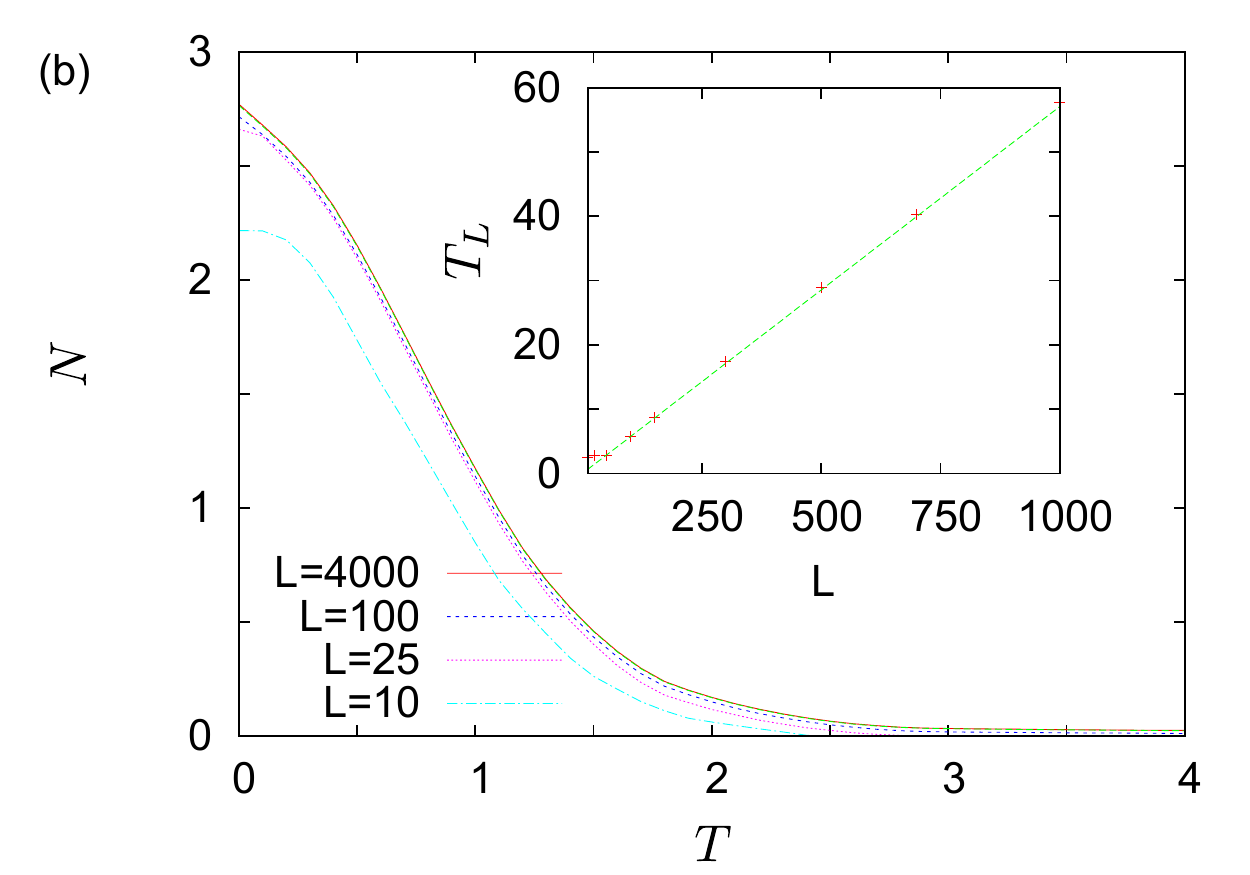}
\caption{(a) Mutual information $I$ per length for the interacting system as a function of temperature for $L=200$, $\kappa=v_f=1$, and $g_0=h_0=f_0=1$. The inset shows the von Neumann entropy $S_\mathrm{el}$ of the reduced density matrix of the electrons. The horizontal lines denote the maximal possible entropies in a system with electron density $\rho_\mathrm{el}=1$ and with different lengths of the linear part of the spectrum, given in per cent of the spectrum, as explained further in the text. (b) Logarithmic negativity $\mathcal{N}$ {per length} as a function of temperature for $\kappa=v_f=1$, $g_0=0.1$, and various system sizes. The inset shows the finite-size dependence of the characteristic temperature $T_L$ at which $\mathcal{N}$ vanishes for $\alpha=1.0$.
\label{fig:int_mutual_information_neg}}
\end{figure}

Results for the mutual information are shown in Fig.\ \ref{fig:int_mutual_information_neg}(a). The mutual information first decreases as the temperature is increased from small values, reaches a minimum, and then increases again, approaching a nonzero constant for high temperatures.

To address the range of validity of our description, we plot in the inset of Fig.\ \ref{fig:int_mutual_information_neg}(a) the von Neumann entropy of the reduced density matrix of the electrons. While our calculations are exact for the investigated model, the model is artificial in that the electronic bands are strictly linear over all momenta and energies. A real material is expected to show a linear spectrum close to the Fermi energy but deviations from linearity away from it. In this case, one can still describe the low-energy excitations within Luttinger-liquid theory but results become unphysical beyond this regime. Let us consider a system with $N_\mathrm{tot}$ single-particle states, of which $N_\mathrm{lin}<N_\mathrm{tot}$ belong to the linear part of the spectrum. The maximal von Neumann entropy per length of the reduced density matrix of the states in the linear spectrum is
\begin{align}
\frac{S^\mathrm{max}_\mathrm{Neumann}}{L}
&= \frac{S^\mathrm{MAX}_\mathrm{Neumann}}{N_\mathrm{tot}/\rho_\mathrm{el}}
  = \frac{\rho_\mathrm{el}}{N_\mathrm{tot}} \ln \begin{pmatrix}
 N_\mathrm{lin} \\
 N_\mathrm{lin}/2
 \end{pmatrix} \nonumber \\
&= \ln(2)\, \rho_\mathrm{el}\, \frac{N_\mathrm{lin}}{N_\mathrm{tot}} ,
\end{align}  
where $\rho_\mathrm{el}$ is the real-space electron concentration and the argument of the logarithm in the first line is a binomial coefficient. Comparing the von Neumann entropy $S_\mathrm{el}$ of the reduced density matrix of the electrons to this limit, one can check the validity of the Luttinger-liquid description. The limits are shown in the inset of Fig.\ \ref{fig:int_mutual_information_neg}(a). Comparing with the main panel, we conclude that the initial decrease and the minimum of the mutual information are correctly described since in this regime $S_\mathrm{el}$ is still far below the entropic bound. The high-temperature plateau may be also reached but in realistic systems with finite bandwidth this plateau is truncated when $S_\mathrm{el}$ reaches the bound.

Figure \ref{fig:int_mutual_information_neg}(b) shows the entanglement negativity. The negativity is extensive, $N \sim L$, for large $L$. It decreases with increasing temperature and in a finite system becomes exactly zero above a characteristic temperature $T_L$. The temperature $T_L$ is connected to the smallest wave number $2 \pi / L$ in the system. By expanding Eq.\ (\ref{eq:ent_res}) for small $k$, we obtain
\begin{equation}
T_L = \frac{5}{4 \pi (2 \pi)^{\alpha} \ln 2}\, h_0\, L^{\alpha} .
\end{equation}
Hence, for $\alpha=0$ the characteristic temperature $T_L$ is independent of the system length $L$. For less strongly screened interaction, i.e., $\alpha>0$, the characteristic temperature grows with the system size and diverges in the thermodynamic limit.

\section{Conclusions and outlook}
\label{sec:conc}

To summarize, we have addressed correlations and entanglement between the electronic (charge) and lattice degrees of freedom of a one-dimensional chain. We have calculated the entanglement entropy at zero temperature and the mutual information and entanglement negativity at finite temperatures. This was done for two models, on the one hand the original Wentzel-Bardeen model without electron-electron interaction and with electron-phonon coupling linear in momentum and on the other a generalized Wentzel-Bardeen-Luttinger model with electron-electron interaction and electron-phonon coupling scaling as the square root of momentum.

As noted above, the entanglement entropy of a coupled oscillator system follows an area law, whereas the entanglement entropy of a fermionic system generally follows an area law with log corrections.
However, if we consider these systems in their most natural physical realizations, i.e., the phonon and electron subsystems of a solid, the situation may change. We have found that the entanglement entropy between a subset of the lattice of length $l<L$ and the rest of the system shows volume-law scaling, $S_l\sim l$, due to the interaction between the electron and lattice subsystems. Consequently, the entanglement entropy between the whole lattice and the electrons is proportional to the full length, $S \sim L$. We suggest that a similar phenomenon may occur for trapped cold atoms in an optical resonator, where the photons play the role of the coupled bosons.

To check the validity of the Luttinger description for real systems with finite bandwidth, we have evaluated the von Nemuann entropy of the reduced density matrix of the electrons (at $T=0$ this is the entanglement entropy). This electronic entropy must satisfy an upper bound, the violation of which signals the breakdown of our description. We find that at temperatures corresponding to thermal energies low compared to the electronic bandwidth our description is valid.

Both the mutual information and the negativity initially decrease with increasing temperature. The mutual information goes through a minimum and approaches a finite constant for $T \to \infty$. For sufficiently large electronic bandwidth, this plateau can still exist but we conjecture that it is cut off at high temperatures when the electronic entropy starts to violate the aforementioned bound. The negatively monotonously decreases with increasing temperature and become zero above a characteristic temperature, which implies that the entanglement disappears \cite{werner2001}. The characteristic temperature increases with system size if the electron-electron interaction is unscreened or weakly screened. However, for sufficiently strong screening the characteristic temperature remains finite in the thermodynamic limit, indicating the presence of a real phase transition with the entanglement negatively acting as its order parameter.

For the artificial model with infinite bandwidth, the entanglement entropy diverges for sufficiently strong electron-phonon coupling at the Wentzel-Bardeen singularity. The Luttinger description is expected to break down as this singularity is approached. We conjecture that in systems with finite bandwidth the entanglement entropy per length reaches a maximum instead of diverging. This is an interesting issue for future studies.


\begin{acknowledgments}
C.~T. acknowledges financial support by the Deutsche Forschungsgemeinschaft, in part through Collaborative Research Center SFB 1143, project A4, and the Cluster of Excellence on Complexity and Topology in Quantum Matter ct.qmat (EXC 2147).
\end{acknowledgments}

\end{document}